# Spin Transistor and Quantum Spin Hall Effects in $CdB_xF_{2-x}$ - $p$–$CdF_2$ – $CdB_xF_{2-x}$ Sandwich Nanostructures

N.T. Bagraev,[a,*] O.N. Guimbitskaya,[b] L.E. Klyachkin,[a] A.A. Kudryavtsev,[a]

A.M. Malyarenko,[a] V.V. Romanov,[b] A.I. Ryskin,[b] I.A. Shelykh,[b] A.S. Shcheulin[b]

[a]Ioffe Physical-Technical Institute RAS, St. Petersburg, Russia

[b]St.Petersburg Polytechnical University, 195251 St.Petersburg, Russia

**Abstract**

Planar $CdB_xF_{2-x}$ - $p$–$CdF_2$ – $CdB_xF_{2-x}$ sandwich nanostructures prepared on the surface of the $n$-type $CdF_2$ bulk crystal are studied to register the spin transistor and quantum spin Hall effects. The current-voltage characteristics of the ultra-shallow $p^+$-$n$ junctions verify the $CdF_2$ gap, 7.8 eV, and the quantum subbands of the 2D holes in the $p$–type $CdF_2$ quantum well confined by the $CdB_xF_{2-x}$ δ-barriers. The temperature and magnetic field dependencies of the resistance, specific heat and magnetic susceptibility demonstrate the high temperature superconductor properties for the $CdB_xF_{2-x}$ δ-barriers. The value of the superconductor energy gap, $2\Delta=102.06$ meV, determined by the tunneling spectroscopy method appears to be in a good agreement with the relationship between the zero-resistance supercurrent in superconductor state and the conductance in normal state, $\pi\Delta/e$, at the energies of the 2D hole subbands. The results obtained are evidence of the important role of the multiple Andreev reflections in the creation of the high spin polarization of the 2D holes in the edged channels of the sandwich device. The high spin hole polarization in the edged channels is shown to identify the mechanism of the spin transistor and quantum spin Hall effects induced by varying the top gate voltage, which is revealed by the first observation of the Hall quantum conductance staircase.

[*] Corresponding author. Tel.: +7-812-292-73-15; fax: +7-812-297-10-17; e-mail: impurity.dipole@mail.ioffe.ru.

The spin analogues of the Hall-effect in semiconductor nanostructures are of great interest in the last decade [1]. The most versions of the spin Hall-effect are based on the replacement of the external magnetic field by the top gate voltage that allows the manipulations of the spin-polarized carriers in quantum wells [2]. However, the experimental verification of the spin Hall-emf is rather difficult because of non-equilibrium in the system of the spin-polarized carriers created by optical pumping [2] or injected from the metal contacts [3]. Therefore the quantum spin Hall-effect that results from the spin-polarized edge channels in a similar way as the well-known quantum Hall-effect has received much attention [4]. The goal of this work is to use the edge channels in the superconductor − semiconductor − superconductor sandwich nanostructures to observe the quantum spin Hall-effect by varying the top gate voltage (Fig. 1). One of the best candidate on the role of a sandwich nanostructure appears to be the $p$-type $CdF_2$ quantum well confined by the $CdB_xF_{2-x}$ δ-barriers that was prepared within Hall geometry on the surface of the $n$-type $CdF_2$ bulk crystal (Fig. 2) [5]. The planar $p^+$-$n$ $CdF_2$ junction was furnished with a top gate that enabled the measurements for the forward and reverse I-V characteristics, tunnelling spectroscopy studies of the subbands of the 2D holes in the $p$-$CdF_2$ – QW and the superconductor energy gap in the $CdB_xF_{2-x}$ δ-barriers as well as the creation the multiple Andreev reflections (MARs). Besides the identification of the $CdF_2$ gap, 7.8 eV, and the energies of the 2D subbands, the relationship between the zero-resistance supercurrent in superconductor state and the conductance in normal state at the energies of the 2D hole subbands appears to be equal to $\pi\Delta/e$ for each pair of peaks (see Figs. 3 a, b and c)). These data are in a good agreement with the value of the superconductor energy gap in the $CdB_xF_{2-x}$ δ-barriers, $2\Delta=102.06$ meV, determined by the tunneling spectroscopy method [5].

Thus, the part played by the 2D subbands in the quantization of the zero-resistance supercurrent and the proximity effect due to MARs [6-8] has emerged from the measurements of the $p$-CdF$_2$ – QW confined by the superconductor CdB$_x$F$_{2-x}$ δ-barriers [5]. The high temperature superconductor properties for the CdB$_x$F$_{2-x}$ δ-barriers were also verified by studying the temperature and magnetic field dependencies of the resistance (Fig. 4), specific heat and magnetic susceptibility [5]. As was to be expected, the application of the external magnetic field results in the shift in the resistance drop to lower temperatures (Fig. 4), with the value corresponding to the drop of the diamagnetic response [5]. Furthermore, the oscillations of the magnetic susceptibility were revealed by varying both the magnetic field and the temperature value, which seem to result from the vortex manipulation in the nanostructured CdB$_x$F$_{2-x}$ δ-barriers [5]. The results obtained show that the high temperature superconductivity of these sandwich structures seems to be caused by the transfer of the small hole bipolarons through the negative-U dipole centers of boron, B$^+$-B$^-$. The same trigonal dipole boron centers inserted in the superconductor δ-barriers have been found also in the silicon sandwich structures [9,10]. The advantage of the sandwich structures studied over the semiconductor nanostructures is related to the creation of the edged channels because of the suppression of the superconductivity along the perimeter of the δ-barriers. Thus, the 2D holes are virtually absent in the $p$-CdF$_2$ – QW plane once involved in the sustenance of the superconductor properties of the CdB$_x$F$_{2-x}$ δ-barriers. Therefore only MARs seem to be favourable for the spin-polarized transport of holes along the edge channels, with the spin polarization defined by trigonal symmetry of the boron dipole centers (Fig. 1). The longitudinal conductance of these edged channels demonstrates the oscillations at low values of the top gate voltage that provides the absence of the changes in the sheet density of holes (Fig. 5). These data are evidence of the spin transistor effect, because the amplitude of the conductance oscillations is close to the value of $e^2/h$ and asymmetry with respect to the sign reversal in the top gate voltage appears. These facts point out at the top gate control of the value of the Bychkov - Rashba spin-orbit interaction (SOI) as the principal mechanism of the spin transistor effect [11,12]. It should be noted that both the high temperature superconductor properties of the $CdB_xF_{2-x}$ δ-barriers and the energy spectrum of the 2D hole subbands in the *p*-$CdF_2$ – QW allow the observations of the ballistic transport in the edged channels at high temperatures. Besides, the findings of the spin transistor effect reveal the spin projection on the y – direction, because only this spin orientation results in the maximum spin transistor effect within frameworks of the Bychkov – Rashba SOI mechanism [13]. The spin-polarized holes in the edged channels appear to give rise to the quantum spin Hall effect that is revealed by measuring the $G_{xy}$ plateau at low values of the top gate voltages (Figs. 6 a, b and c) [3]. In addition to this $G_{xy}$ plateau, the Hall quantum conductance staircase was observed for the first time, with the amplitude of the quantum step equal to the value of $e^2/h$ (Fig. 7). These results identify also the principal contribution to the spin polarization in the edged channels from the MARs that are seen to be observable at the lowest values of the top gate voltage (Fig. 8). Finally, the spin orientation of holes in the edged channels following by one of the equivalent positions of the trigonal dipole centers of boron, which is selected by straining is revealed also by the MARs versus the drain-source voltage (Figs. 9 a and b).

Thus, the results obtained are evidence of the important role of the multiple Andreev reflections in the creation of the high spin polarization of the 2D holes in the edged channels of the planar $CdB_xF_{2-x}$ - $p$–$CdF_2$ − $CdB_xF_{2-x}$ sandwich device.

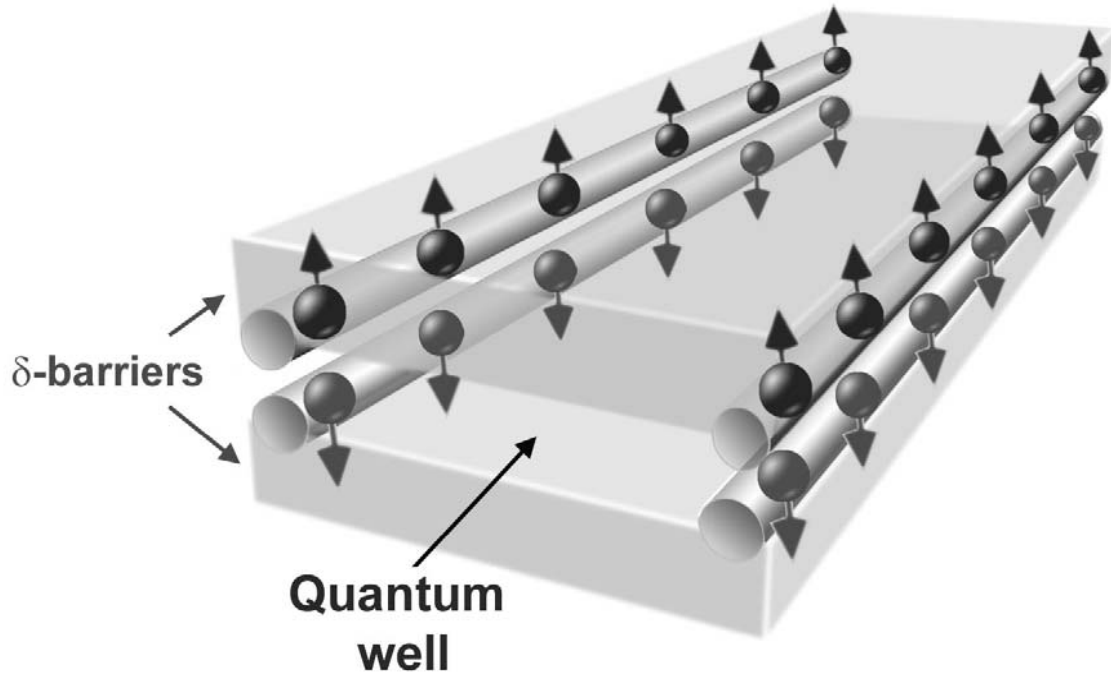

Fig. 1. The scheme of the spin-polarized edge channels in the $CdB_xF_{2-x}/p$-$CdF_2$-QW/$CdB_xF_{2-x}$ sandwich structure that is used in the quantum spin Hall measurements.

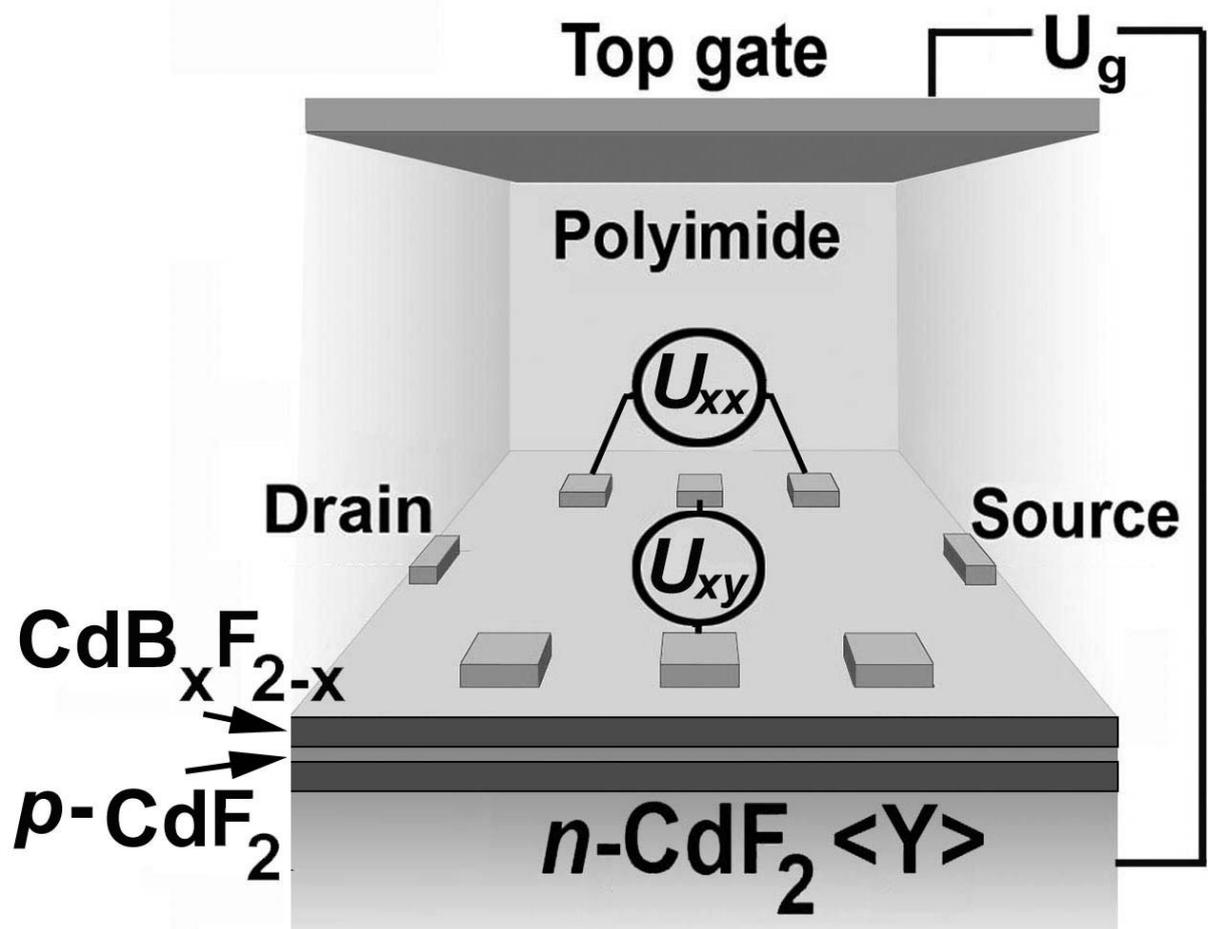

Fig. 2. The scheme of the $p$-type $CdF_2$-QW confined by the δ-barriers of $CdB_xF_{2-x}$ that is prepared on the surface of the $n$-$CdF_2$ crystal, which is provided by the top gate to vary the values of the sheet density of 2D holes and the Rashba spin-orbit interaction as well as to cause the multiple Andreev reflections (MAR).

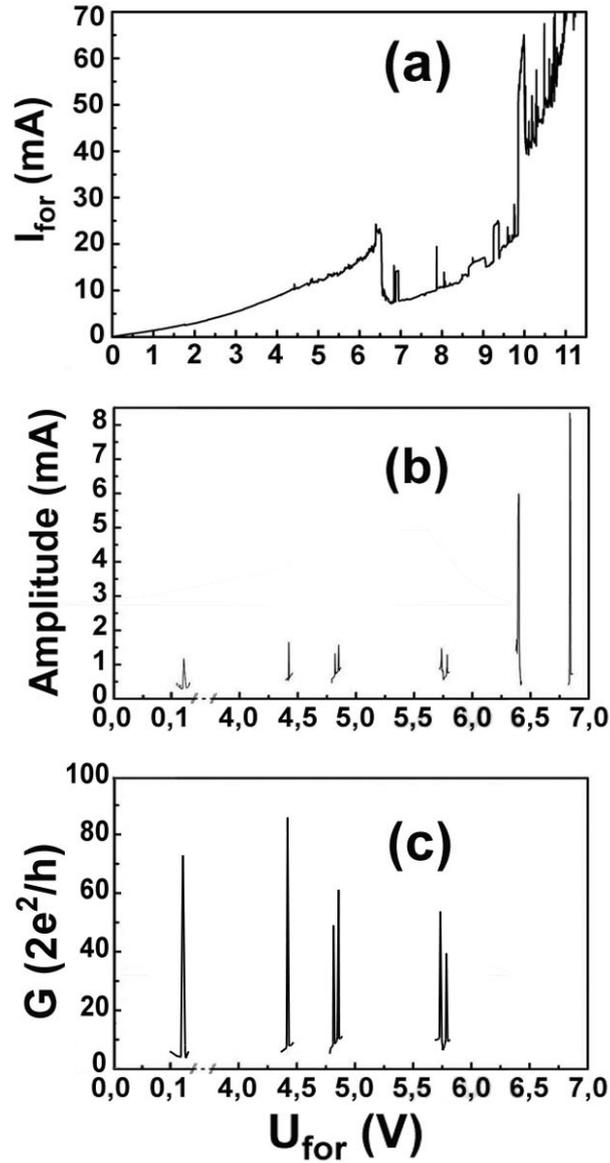

Fig. 3. (a) The forward current-voltage characteristic revealed by the CdB$_x$F$_{2-x}$/$p^+$CdF$_2$-QW/CdB$_x$F$_{2-x}$-$n$-CdF$_2$ junction that reveals the gap value, 7.8 eV, and the valence band structure of the CdF$_2$ crystal. T=300 K.

The spectra of the tunneling forward current (b) and conductance (c) recorded respectively at T=298K and T=345K that identify the energy positions of the subbands of 2D holes in the $p$-type CdF$_2$-QW. T=300 K. The ratio of the magnitude for each current peak and the magnitude for the corresponding conductance peak appears to be equal to $\pi\Delta/e$, where $\Delta$=51.03 meV.

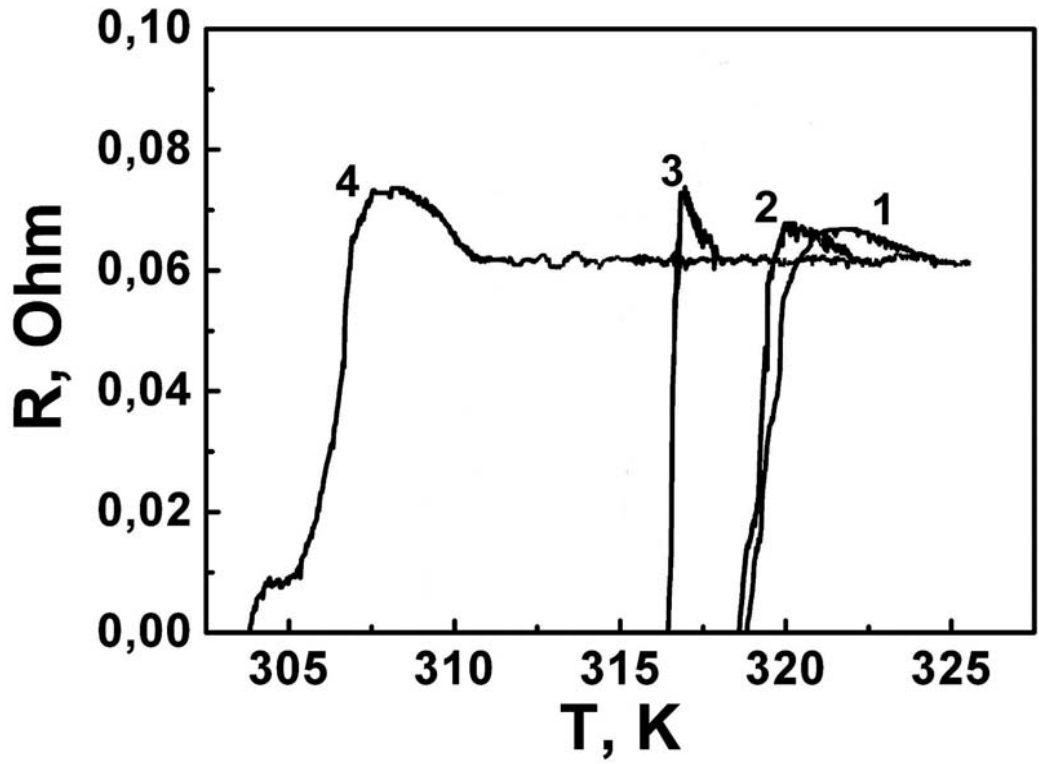

Fig. 4. The temperature dependences of the resistance that were observed in the sandwich structure that represents the *p*-type $CdF_2$-QW confined by the δ-barriers of $CdB_xF_{2-x}$ which is prepared on the surface of the *n*-$CdF_2$ crystal. B, mT: 1 – 0; 2 – 20; 3 – 50; 4 – 100.

1010

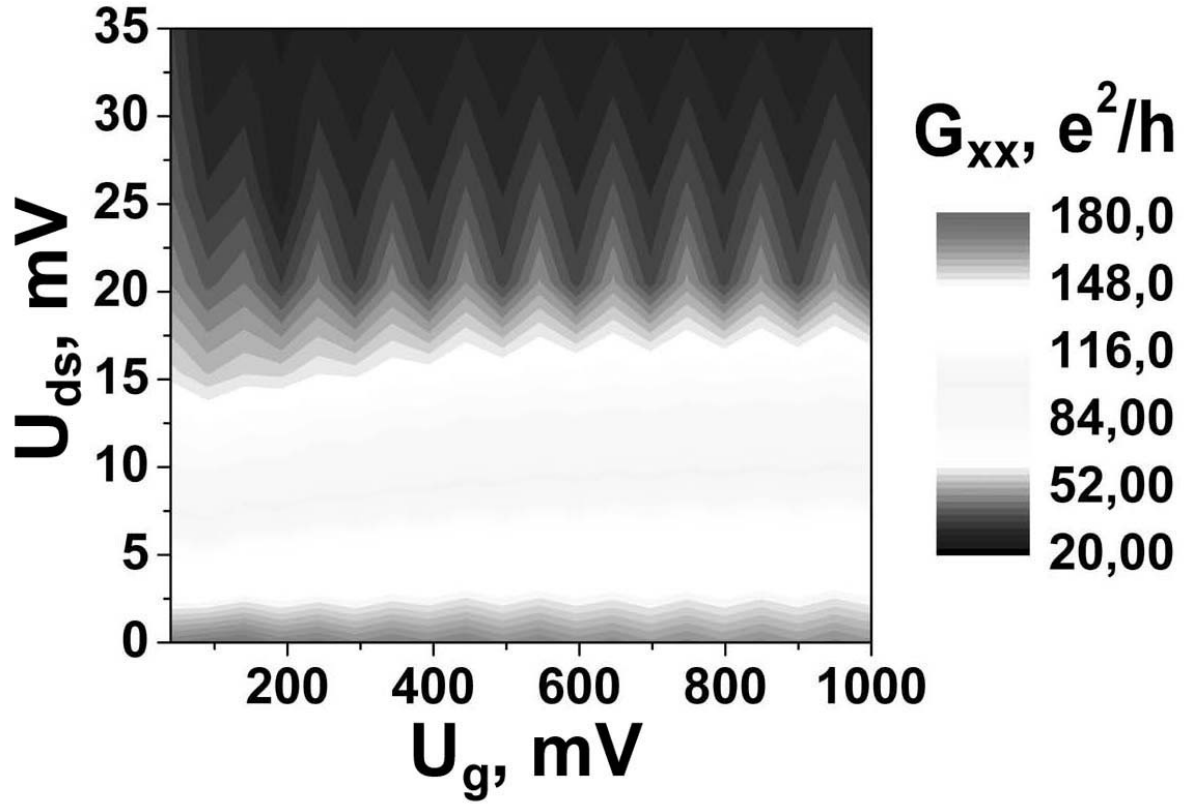

Fig. 5. The conductance oscillations observed in the studies of the $CdB_xF_{2-x}/p$-$CdF_2$-QW/$CdB_xF_{2-x}$ sandwich structure prepared on the surface of the $n$-$CdF_2$ crystal that exhibit the spin transistor effect at T=300 K by varying the top gate voltage. The drain-source current was stabilized, $I_{ds}$ = 10 nA.

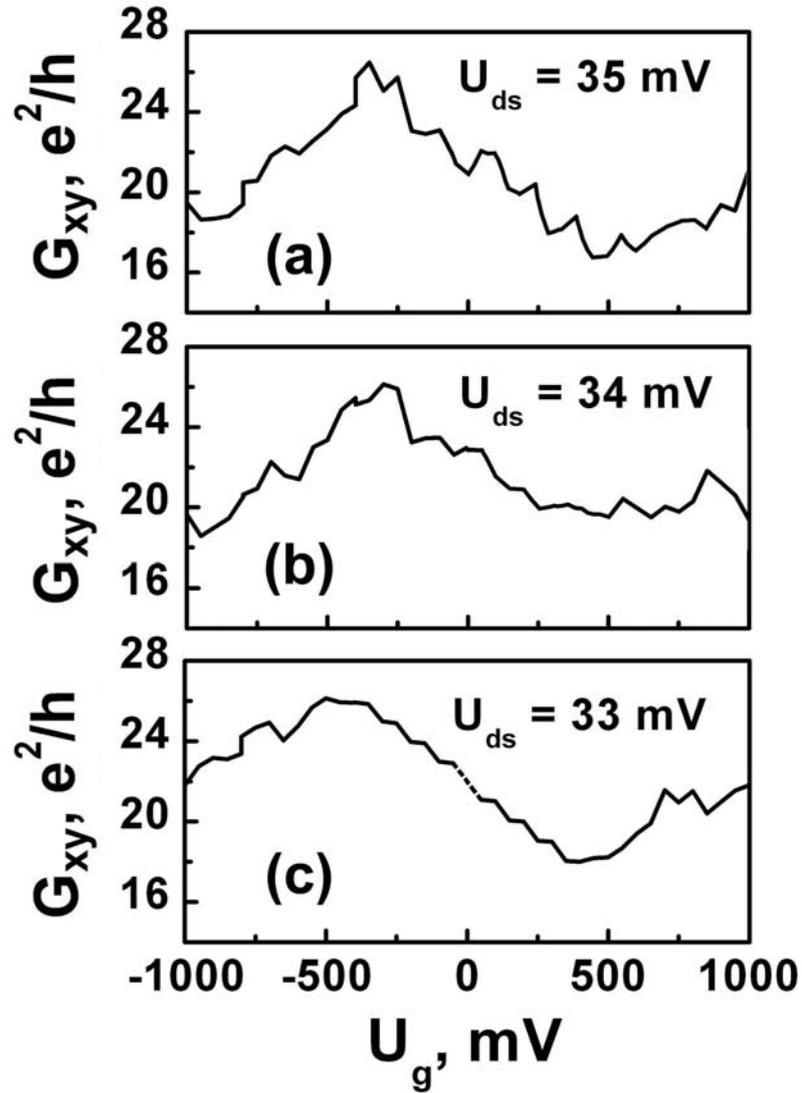

Fig. 6. The Hall conductance revealed by varying the top gate voltage applied to the $CdB_xF_{2-x}/p-CdF_2-QW/CdB_xF_{2-x}$ sandwich structure that is prepared on the surface of the $n-CdF_2$ crystal. $I_{ds}$ = 10 nA. T=300K. The zero-field, $U_g = 0$, conductance value seems to be due to the different spin polarization of holes in the edged channels on the opposite sides of the device(see Figs. 1 and 4).

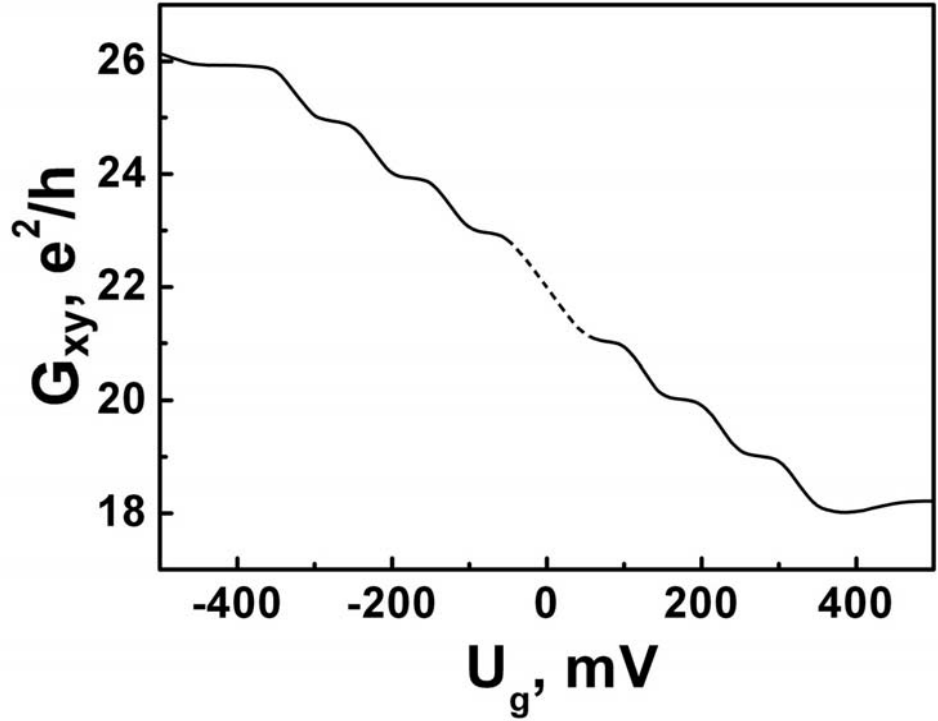

Fig. 7. The Hall quantum conductance staircase that appears to result from the creation of the spin-polarized Hall channels in the $CdB_xF_{2-x}$/*p*-$CdF_2$-QW/$CdB_xF_{2-x}$ sandwich structure by varying the value of the top gate voltage. $I_{ds}$ = 10 nA. T=300K. The quantum step value, $e^2/h$, appears to identify the spin polarization of holes.

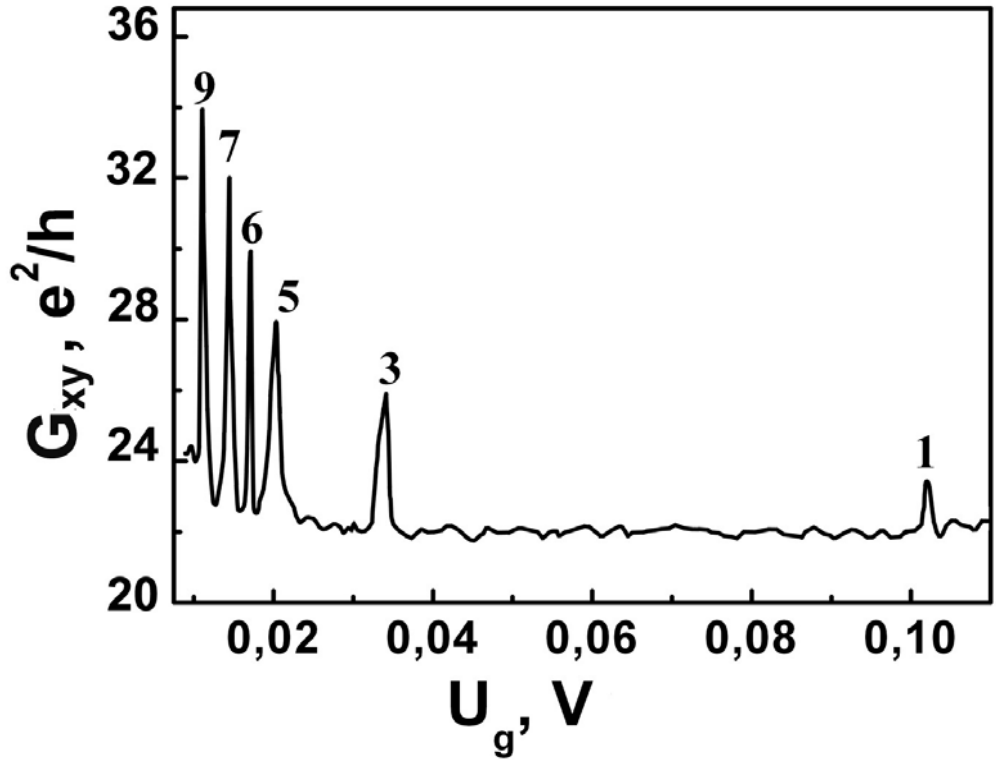

Fig. 8. Multiple Andreev reflections (MARs) derived from the tunneling forward current as a function of the top gate voltage applied to the $CdB_xF_{2-x}/p\text{-}CdF_2\text{-}QW/CdB_xF_{2-x}$ sandwich structure. The MARs peak positions are marked at $V_n = 2\Delta/ne$ with values $n$ indicated. The superconducting gap peak, $2\Delta=102.06$ meV, is also revealed. T=300 K.

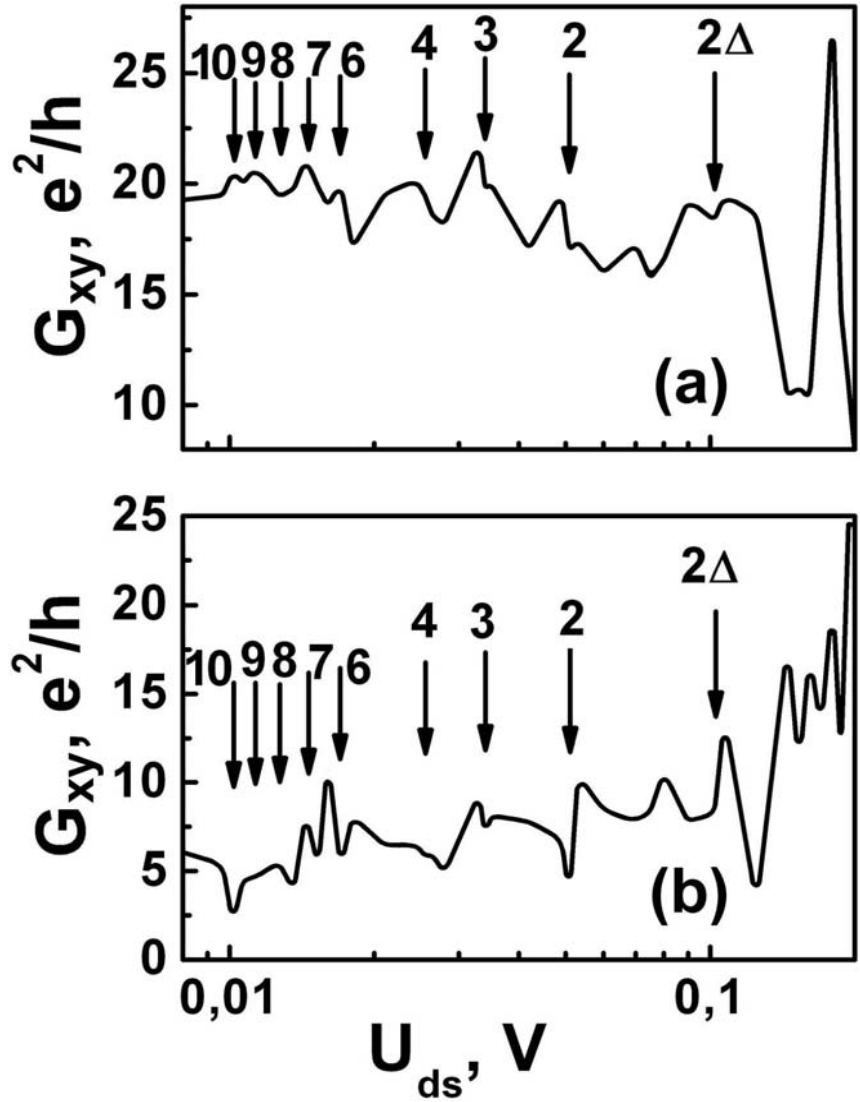

Fig. 9. Multiple Andreev reflections (MARs) that are observed in the dependences of the zero-field, $U_g = 0$, conductance value (a) and the magnitude of the Hall conductance (b) on the drain-source voltage applied to the $CdB_xF_{2-x}/p\text{-}CdF_2\text{-}QW/CdB_xF_{2-x}$ sandwich structure. The MARs peak positions are marked at $V_n = 2\Delta/ne$ with values $n$ indicated. T=300K.